\documentclass[12pt]{article}
\usepackage{putex}
\usepackage{graphicx}
\usepackage{latexsym,amsmath,amsfonts,amssymb}
\usepackage{bbm}
\usepackage{epsfig}
\usepackage{pstricks}
\usepackage{pst-node}

\def\cD{\mathcal{D}}
\def\cF{\mathcal{F}}
\def\cH{\mathcal{H}}
\def\cI{\mathcal{I}}

\def\cN{\mathcal{N}}

\def\cQ{\mathcal{Q}}
\def\cR{\mathcal{R}}
\def\cS{\mathcal{S}}
\def\cT{\mathcal{T}}

\def\bC{\mathbb{C}}

\def\bZ{\mathbb{Z}}

\let\hat\widehat

\DeclareMathOperator{\Tr}{Tr}
\DeclareMathOperator{\Det}{Det}

\def\repa{\raise4pt\hbox{$\square$}\mkern-14mu\raise-4pt\hbox{$\square$}}
\def\repab{\overline{\raise4pt\hbox{$\square$}\mkern-14mu\raise-4pt\hbox{$\square$}\mkern-1mu}}

\def\ie{\textit{i.e.}}

\newcommand{\scF}{\ensuremath{\mathcal{F}}}
\newcommand{\scG}{\ensuremath{\mathcal{G}}}
\newcommand{\scH}{\ensuremath{\mathcal{H}}}
\newcommand{\scI}{\ensuremath{\mathcal{I}}}

\newcommand{\scN}{\ensuremath{\mathcal{N}}}

\newcommand{\scR}{\ensuremath{\mathcal{R}}}

\newcommand{\scV}{\ensuremath{\mathcal{V}}}
\newcommand{\scW}{\ensuremath{\mathcal{W}}}

\newcommand{\be}{\begin{equation}}
\newcommand{\ee}{\end{equation}}
\newcommand{\bea}{\begin{equation}\begin{aligned}}
\newcommand{\eea}{\end{aligned}\end{equation}}
\newcommand{\beq}{\begin{eqnarray}}
\newcommand{\eeq}{\end{eqnarray}}

\newcommand{\mmod}[1]{[\![  #1 ] \! ]}

\newcommand{\unit}{\mathbbm{1}}

\begin{document}

\title{4d Index to 3d Index and 2d TQFT}

\authors{Francesco Benini$^\diamondsuit$, Tatsuma Nishioka$^\diamondsuit$, and Masahito Yamazaki$^\spadesuit$}

\institution{PU}{
${}^\diamondsuit$Department of Physics, Princeton University, Princeton, NJ 08544, USA}
\institution{PCTS}{${}^\spadesuit$Princeton Center for Theoretical Science, Princeton University, NJ 08544, USA}

\abstract{We compute the 4d superconformal index for
$\cN=1, 2$ gauge theories on $S^1\times L(p,1)$, where
$L(p,1)$ is a lens space. We find that
the 4d $\scN=1, 2$ index on $S^1\times L(p,1)$ reduces to
a 3d $\cN=2,4$ index on $S^1\times S^2$ in the large $p$
limit, and to a 3d partition function
on a squashed $L(p,1)$ when the size of the temporal $S^1$ shrinks to zero.
As an application of our index, we study 4d $\scN=2$ superconformal
field theories arising from the 6d $\cN=(2,0)$ $A_1$ theory on a
punctured Riemann surface $\Sigma$,
and conjecture the existence of a 2d Topological Quantum Field Theory on $\Sigma$ whose correlation
function coincides with the 4d $\cN =2$ index on $S^1 \times L(p,1)$.
}

\preprint{PUPT-2390}

\maketitle


\tableofcontents

\section{Introduction}
\label{sec.intro}

One of the beauties of supersymmetric gauge theories is that they are
often amenable to exact analysis. Recent studies have uncovered powerful
techniques (mostly based on localization) to extract exact results for 3d and 4d supersymmetric gauge
theories, including the 4d $\scN\ge 1$ superconformal index
on $S^1\times S^3$ \cite{Romelsberger:2005eg,Kinney:2005ej},
the 4d $\scN\ge 2$ partition function on $S^4$ \cite{Pestun:2007rz}, the 3d $\scN\ge 2$ partition
function on $S^3$ \cite{Kapustin:2009kz,Drukker:2010nc,Herzog:2010hf,Jafferis:2010un,Hama:2010av}, and the 3d $\scN\ge 2$ index on
$S^1\times S^2$ \cite{Bhattacharya:2008zy,Kim:2009wb,Imamura:2011su}.

Given the richness of the subject, a natural question is whether
there are precise relations among different quantities.
One such relation has been noticed by
\cite{Dolan:2011rp,Gadde:2011ia,Imamura:2011uw} (see also \cite{Nishioka:2011dq}),
which shows that a 4d index
on $S^1\times S^3$ reduces to a
3d partition function on $S^3$
when the radius of the temporal $S^1$ goes to zero.
We will present yet another connection between 4d and 3d quantities.

In this paper we study the superconformal index of 4d $\scN=1,2$
superconformal field theories (SCFTs) on $S^1\times L(p,1)$, and
obtain explicit expressions for them.\footnote{The index of $\cN = 4$
super-Yang-Mills on $S^1 \times S^3/\bZ_p$ was
studied in \cite{Lin:2005nh,Hikida:2006qb}.}
This is the first result of our paper, see section \ref{sec.4dindex}
and in particular the expressions in \eqref{N2I}--\eqref{chemical} and \eqref{N1B0}--\eqref{N1I0} for
the result and the appendix for the derivation.
Here $L(p,q)$, where $p,q$ are coprime integers, is the lens space defined as the orbifold of
$S^3: \{(z_1, z_2)\in \bC^2 \,\big|\, |z_1|^2+|z_2|^2=1\}$ under
the identification
\beq
(z_1, z_2)\sim \left(e^{2\pi i q/p} z_1, \, e^{-2\pi i /p} z_2\right) \ ,
\label{lenseorbifold}
\eeq
where $SU(2)_1$ acts on $(z_1,z_2)$ as a doublet (see section 2 for our notation).
Without loss of generality one can assume $0<p$ and $0 < q \le p-1$.
As for fermions, we choose the orbifold action such that
the supercharges $\bar\cQ_{I\dot\alpha}$ are preserved, while
$\cQ^I_\alpha$
are broken.
Note that this action has no fixed points, and the manifold $L(p,q)$ is still smooth.
In this paper%
\footnote{With respect to v1, in v2 of this paper the discussion of the spaces $L(p,q)$ with $q\neq 1,p-1$ has been dropped because it was valid only for very special theories. We thank L.F. Alday and J. Sparks for pointing this out to us.}
we consider the case $q=1$:
$L(p,1)$ is the orbifold $S^3/\bZ_p$, where $\bZ_p$ acts on
the $S^1$ fiber of the Hopf fibration.
Equivalently, the $\bZ_p$ action is embedded into $U(1)_1\subset
SU(2)_1$.

Our $S^1\times L(p,1)$ index in itself will serve as a useful
tool to quantitatively study the strongly
coupled IR fixed points. For example our index could be used for checks of
4d $\scN=1$ Seiberg dualities. Mathematically, such a duality is expressed as an
identity involving an integral of a generalization of the elliptic Gamma
function.%
\footnote{For the generalization of the elliptic Gamma function see the infinite product form in \eqref{prod1} or \eqref{prod2}, while for its hyperbolic version see (\ref{3dsquashed1loop}).}

The second result is about the compactification of the 4d theories.
When 4d $\scN=1,2$ SCFTs are compactified on $S^1$, they flow in
the IR to 3d $\scN=2,4$ SCFTs.
In our setup we have two circles:
one circle (denoted by $S^1_{T}$) is the temporal $S^1$,
and another (denoted by $S^1_{H}$) is the $S^1$ of the Hopf
fibration.
Depending on the choice of $S^1$, we can study two limits of
our index.%
\footnote{It is important to keep in mind that
the actual meanings of $S^1\to 0$ are different in
the two cases. In the limit $S^1_H \to 0$ we consider the orbifold $\bZ_p$
in the limit $p\to \infty$,
whereas in the limit $S^1_T\to 0$ we shrink the
size of $S^1_T$ without taking an orbifold. In the former case
the KK modes along $S^1_H$ are projected out in the orbifolding process
(section \ref{sec.3dindex}),
while in the latter case the KK modes decouple from the constant modes
but still remain, and we will have to subtract the divergent part (section \ref{sec.3dPF}) as explained later.}

In the limit $S^1_{H}\to 0$, {\it i.e.} $p\to \infty$,
the lens space $L(p,1)$ reduces to the two sphere and we
show that the 4d index reduces to the 3d index on $S^1_T\times S^2$ (section
\ref{sec.3dindex}):
\be
\cI^{\rm 4d}[S^1_T \times L(p,1)] \quad\xrightarrow{p\to\infty}\quad \cI^{\rm 3d}[S^1_T\times S^2] \;.
\ee
In this limit the holonomies of the gauge field along $S^1_H$ in the 4d theory are mapped to the
monopole charges in the 3d theory.
On the other hand in the limit $S^1_{T}\to 0$ the temporal circle shrinks to zero and the 4d index reduces to the 3d partition function on  $L(p,1)$ (section \ref{sec.3dPF}):
\be
\cI^{\rm 4d}[S^1_T\times L(p,1)] \quad \to \quad Z^{\rm 3d}[L(p,1)]  \qquad\qquad \text{when} \quad S^1_{T} \to 0 \;.
\ee

\begin{figure}[t]
\centering
\scalebox{0.4}{\input{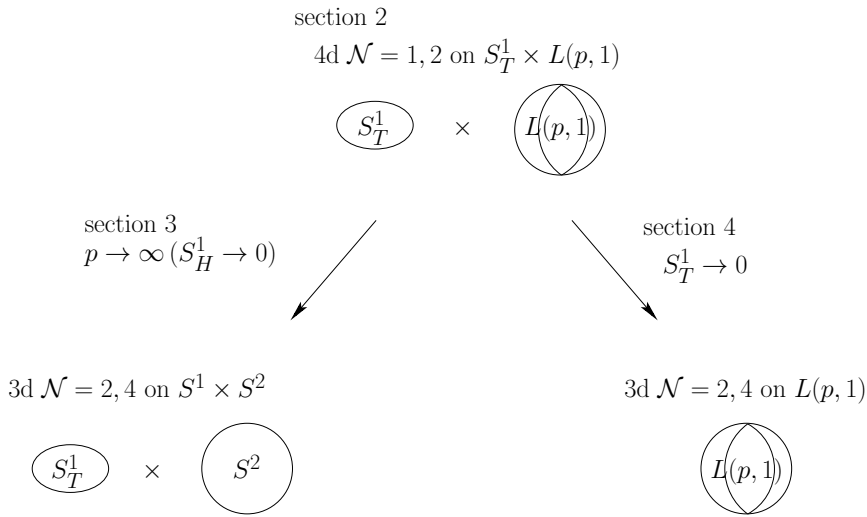}}
\caption{A schematic summary of the relations obtained in section 2, 3 and 4.}
\label{fig.map}
\end{figure}

The third result is about an application of our index (section \ref{sec.2dTQFT}).
When the 4d theory arises from the 6d $(2,0)$ theory on a
punctured Riemann surface $\Sigma$,
we conjecture the existence of a 2d topological quantum field theory (TQFT) on $\Sigma$ whose correlation
function coincides with the 4d index on $S^1\times L(p,1)$,
generalizing a similar claim of \cite{Gadde:2009kb,Gadde:2011ik}.
We summarize the relations between the 4d index and the 3d quantities
we will obtain in this paper in figure \ref{fig.map}.

\section{4d Index on $S^1 \times L(p,1)$}
\label{sec.4dindex}

In this section we will present our expression for the 4d $\cN =1, 2$
superconformal indices\footnote{Despite the name, we can define this index for 4d $\scN=1, 2$ theories
which are non-conformal in the UV.} on
$S^1 \times L(p,1)$. The derivation of these
results
is given in appendix \ref{app: derivation}.

Consider a 4d $\scN=1$ ($\scN=2$) superconformal field theory on
$S^1\times S^3$. Its superconformal algebra is given by $SU(2,2|1)$
($SU(2,2|2)$). We let the R-symmetry index and the supercharges be $I=1$
($I=1,2$) and
$\cQ^I_\alpha$, $\bar \cQ_{I\dot\alpha}$, $\cS_{I\alpha}$,
$\bar\cS^I_{\dot\alpha}$, respectively. Here $\alpha=\pm$ ($\dot\alpha=\pm$)
is the index for the $SU(2)_1$ ($SU(2)_2$) spin  of
the $SO(4)\simeq SU(2)_1 \times SU(2)_2$ rotational symmetry of the three sphere.

The orbifold theory has a set of degenerate vacua, labeled by a non-trivial holonomy
$V$ along the $S^1_H$ direction, since $\pi_1(L(p,1)) = \bZ_p$.
The holonomy $V$ satisfies $V^p=1$ and can be mapped to an element of the maximal torus
through conjugation by an element of the gauge group
\be
\label{holonomy V}
V=(\omega^0 \unit_{N_0}, \cdots , \omega^{p-1} \unit_{N_{p-1}}) \;,
\ee
where $\omega = e^{2\pi i/p}$ and the integers $N_I$ satisfy the relation $\sum_{I=0}^{p-1} N_I = N$ with $N$ the rank of the gauge group.
Another useful parametrization is given by $m_1,\ldots, m_N$, which is defined as
\be\label{holonomies}
(m_i) = (\underbrace{0,\cdots,0}_{N_0},\cdots,\underbrace{p-1,\cdots,p-1}_{N_{p-1}}) \;,\qquad\qquad N_I = \big( \# m_i = I \big) \;.
\ee
where $i=1,\ldots, N$ and $I=0,\cdots,p-1$.
In this notation the $i$-th holonomy is given by $\omega^{m_i}$.
The holonomy breaks the gauge group into a product of $p$ subgroups
\be
G \, \to\,   \prod_{I=0}^{p-1} G_I \ ,
\label{gaugebreaking}
\ee
where the rank of $G_I$ is given by $N_I$.
For example, in case of a $U(N)$ gauge group we have
$U(N) \, \to\,   \prod_{I=0}^{p-1} U(N_I)$.

\subsection{$\scN=2$ Index}

Let us begin with the $\scN=2$ index. We will comment on the $\scN=1$ index later.

We define the index with respect to the supercharge $\cQ \equiv \bar\cQ_{2+}$ that
survives the orbifold projection.
This is given by \cite{Romelsberger:2005eg,Kinney:2005ej}
\be\label{4dindex}
\cI = \Tr \, (-1)^\cF e^{- \tilde\beta \, \Xi} \, t^{2(E+j_2)} \, y^{2j_1} \, v^{-(r+R)} \, z^F \;,
\ee
where $\cF$ is the fermion number, the trace is taken over the states of the theory on $S^3$, and the quantum numbers of the R-symmetries $U(1)_R \subset SU(2)_R$ and $U(1)_r$
are denoted by $(R,r)$. The $\Xi$ is the commutator of $\cQ$ with its conjugate:
\beq
\label{N2QQ}
\Xi \equiv 2\{\cQ,\cQ^\dag\}=E-2j_2-2R+r \;,
\eeq
and the index is independent of $\tilde\beta$. Therefore, only the states obeying $\Xi=0$ contribute to the index. The expression $z^F$ is a shorthand for $\prod_j z_j^{F_j}$, where $F_j$ are charges with respect to flavor symmetries (commuting with $\cQ, \cQ^{\dagger}$) and $z_j$ are their chemical potentials. The operators $E + j_2$, $j_1$, $R+r$ and $F_i$ appearing in \eqref{4dindex} are the maximal set of operators commuting with $\cQ$ and $\cQ^{\dagger}$.

In the path integral formulation, our 4d index for an $\scN=2$ theory on $S^1\times L(p,1)$ can be written as
\be
\label{N2I}
\cI_{p}(t,y,v,z) = \sum_{m} \scI^0_{p,m}(t,y,v,z) \int \big[ da \big] \, \exp\left[ \sum_{n=1}^\infty \frac1n
\, \hat{\scI}_{p,m}(t^n,y^n,v^n,z^n;e^{i n a}) \right] \;.
\ee
Here, the index consists of a sum of indices labeled by the set of holonomies $m\equiv \{m_i\}$,
with $0 \leq m_1 \leq \cdots \leq m_N \leq p-1$.
The measure $\big[ da \big]$ is given by
\be
\big[ da \big] = \frac1{\prod_I |\scW_I|} \prod_{i=1}^N \frac{da_i}{2\pi} \prod_{\substack{\alpha \in G \\ \alpha(m)=0}} 2 \sin \frac{\alpha(a)}{2} \;,
\label{measureWeyl}
\ee
where $\scW_I$ is the Weyl group of $G_I$ and the last product is over
the roots of the unbroken gauge group. This is the Haar measure of the unbroken gauge group $\prod_I G_I$.

The function $\hat{\scI}_{p,m}$ is the single-letter contribution to the
index, and is obtained by summing over all the
fields $\Phi$ contributing to the index:
\beq
\hat{\scI}_{p,m}(t,y,v,z;e^{ia})=\sum_{\Phi} \hat{\scI}_{p, m}^{\Phi}(t,y,v,z;e^{ia}) \ .
\eeq
For a vector multiplet we find (see appendix \ref{app: derivation})
\be
\label{I2vector}
\hat{\scI}^{\scN=2 \textrm{ vector}}_{p,m}(t,y,v,z;e^{ia})=
\sum_{\rho \,\in\, \text{Adj}}
\big[ \left(t^2 v-t^4 v^{-1}+t^6-1 \right)
F_{p}(t,y;\mmod{\rho(m)})+\delta_{\mmod{\rho(m)},0}
\big] \, e^{i \rho(a)}  \ ,
\ee
and for a half-hypermultiplet in a representation $\scR$ and with flavor
charges $F$
\be
\label{I_halfhyper}
\hat{\scI}^{\scN=2 \text{ half-hyper}}_{p,m}(t,y,v,z;e^{ia}) =  \sum_{\rho \,\in\, \cR} \big( t^2 v^{-1/2} z^F - t^4
v^{1/2} z^{-F} \big) \, F_{p}(t,y;\mmod{\rho(m)})  \, e^{i \rho(a)} \;,
\ee
where the summations are over the weights of the adjoint representation and of the representation $\scR$ of $G$ respectively.%
\footnote{For an adjoint
representation, this sum is over the roots of $G$ as well as vanishing weights.}
Here the function $F_{p}(t,y;L)$ is defined by
\bea
F_{p}(t,y;L) = \frac{1}{1-t^{6}}
 \bigg(
\frac{t^{3L}y^{L}}{1-t^{3p}y^p} +
\frac{t^{3(p-L)}y^{-(p-L)}}{1-t^{3p}y^{-p}}
\bigg) \ ,
\label{qne0}
\eea
and we use the notation
\be
\label{mmod definition}
\mmod{x}  = \{\text{ an integer }~ y ~\text{ such that }~ 0 \leq y < p ~\text{ and
}~ y \equiv x \pmod{p}  \} \;.
\ee

Finally $\scI^0_{p,m}(t,y,v,z)$ in front of the integral is the contribution from zero-point oscillations
of the fields, which depends on the matter content of the
theory. We have
\begin{multline}
\label{N2I0}
\scI^0_{p,m}(t,y,v,z) = \exp \Bigg[ \beta \sum_{\alpha\in G} \frac{1+\mu}{2p} \big( p \mmod{\alpha(m)} - \mmod{\alpha(m)}^2 \big) \\
- \beta \sum_{\Phi:\,\text{half-hyper}} \sum_{\rho\in \scR_{\Phi}} \frac{1 + \mu - 2F_\Phi \nu}{4p} \big( p \mmod{\rho(m)} - \mmod{\rho(m)}^2 \big) \Bigg] \;,
\end{multline}
where we introduced the notation
\be
t = e^{-\frac{\beta}{2}} \;,\qquad
y = e^{-\beta \Omega_1} \;,\qquad
v = e^{-\beta \mu} \;\qquad z_j=e^{-\beta \nu_j} \;.
\label{chemical}
\ee
Note that this vanishes when the holonomy is trivial: $m=0$, so that in the first term we only considered the sum over the roots of $G$. Moreover, since $\mmod{-x} = p-\mmod{x}$, we have $p\mmod{x} - \mmod{x}^2 = p\mmod{-x} - \mmod{-x}^2$.

For concreteness, let us specialize to a $U(N)$ gauge theory with vector
multiplets and tri-fundamental hypermultiplets. This is the theory we will
discuss in section \ref{sec.2dTQFT}. We also set the flavor
chemical potential to zero, $z=1$. Then the measure $\big[ da \big]$ becomes
\be
\label{brokenHaarmeasure}
\big[ da \big] = \frac1{\prod_I N_I!} \prod_{i=1}^N \frac{da_i}{2\pi} \prod_{\substack{i,j \\ m_i = m_j}} 2 \sin \frac{a_i - a_j}2 \ ,
\ee
which coincides with the product of Haar measures $\prod_I \big[ dU_I \big]$ with
$N_I\times N_I$ unitary matrices $U_I$, whose eigenvalues are
denoted by $e^{i a_i}$. We have
\bea
\label{simplevector}
\hat{\scI}^{\scN=2 \textrm{ vector}}_{p,m}(t,y,v;e^{ia})&= \sum_{i,j=1}^N f_{p} \big( \mmod{m_i - m_j} \big) \, e^{i(a_i - a_j)}
= \sum_{I,J=0}^{p-1} f_{p} \big( \mmod{I - J} \big) \, \Tr(U_I)
\Tr(U_J^{\dagger})
\eea
for a vector multiplet, and
\bea
\hat{\scI}^{\scN=2 \textrm{ tri-fund}}_{p,m}(t,y,v;e^{ia}) & =\sum_{i,j,k=1}^N g_{p} \big( \mmod{m_i+m_j+m_k} \big) \, e^{i (a_i +a_j+a_k)} \\
& = \sum_{I,J,K=0}^{p-1} g_{p} \big( \mmod{I+J+K} \big) \, \Tr(U_I)
\Tr(U_J) \Tr(U_K)
\label{simpletri}
\eea
for a half-hypermultiplet in the tri-fundamental representation.
Here we defined
\bea
f_{p}(L)&=(t^2 v-t^4 v^{-1}+t^6-1) \, F_{p}(t,y;L)+\delta_{L,0} \\
g_{p}(L)&=(t^2 v^{-\frac{1}{2}}-t^4 v^{\frac{1}{2}}) \, F_{p}(t,y;L) \; .
\label{fgdef}
\eea

\subsection{$\scN=1$ Index}

Let us repeat the discussion for the 4d $\scN=1$ index given by
\be
\cI(t,y,z) = \Tr(-1)^\cF \, t^{2(E+j_2)} \, y^{2j_1} \, z^F \;,
\label{N1index}
\ee
where the trace is taken over all the fields satisfying
\be
\{ \cQ,\cQ^{\dagger} \}=E-2j_2-\frac{3}{2}\tilde r =0 \;,
\label{N1QQ}
\ee
and $\tilde{r}$ is the R-symmetry of the $\scN=1$ superalgebra.
When we regard the $\scN=2$ SCFT as the $\scN=1$ SCFT,
the R-symmetries $R, r$ of the $\scN=2$ SUSY recombine into the $\scN=1$ R-symmetry $\tilde r$ and a flavor symmetry $A$ commuting with $\cQ$.
By comparing the definitions of the indices (compare \eqref{4dindex}-\eqref{N2QQ} with \eqref{N1index}-\eqref{N1QQ}, see also \cite{Shapere:2008zf}) we obtain
\be
\tilde r = \frac{4R - 2r}3 \ , \qquad\qquad A = -R - r \;.
\ee

The same derivation as in  appendix \ref{app: derivation} works for the $\scN=1$ theory,
but there are some important differences.
First, we do not have a chemical potential $v$ for the R-symmetry.
Second, there is a zero-point contribution $e^{i B^0_{p,m}(a)}$ to the measure
of the theory. We have
\be
B^0_{p,m}(a)= - \sum_{\Phi:\,\text{chiral}} \sum_{\rho \in \cR_\Phi} \frac{\rho(a)}{2p} \big( p \mmod{\rho(m)} - \mmod{\rho(m)}^2 \big) \;.
\label{N1B0}
\ee
This correction is absent for a vector-like theory, including the $\scN=2$ theories previously discussed. The index is given by
\be
\label{N1I}
\cI_{p}(t,y,z) = \sum_{m} \scI^0_{p,m}(t,y,z) \int \big[da\big] \, e^{iB^0_{p,m}(a)}\, \exp\left[ \sum_{n=1}^\infty \frac1n \, \hat{\scI}_{p,m}(t^n,y^n,z^n;e^{i n a}) \right] \;.
\ee
The single-letter index $\hat{\cI}$ for an $\cN=1$ vector multiplet is
\be
\label{single letter N=1 vector}
\hat\cI^{\scN=1 \text{ vector}}_{p,m} = \sum_{\rho \in \text{Adj}}
\left( (t^6-1) F_{p}(t,y;\mmod{\rho(m)}) + \delta_{\mmod{\rho(m)},0} \right) \, e^{i\rho(a)}  \;,
\ee
where we used the same function \eqref{qne0}.
This is essentially the half of \eqref{I2vector} corresponding to an $\cN=1$ vector multiplet.
For an $\scN=1$ chiral multiplet with flavor charges $F$ we have
\bea
\label{single letter N=1 chiral}
\hat \cI^{\scN=1 \textrm{ chiral}}_{p,m} &= \sum_{\rho \in \cR} \left(t^{3Q} z^F e^{i\rho(a)}- t^{6-3Q} z^{-F} e^{-i\rho(a)}\right) \, F_{p}\big( t,y;\mmod{\rho(m)} \big) \;.
\eea
In this expression we have included an anomalous R-charge $Q$
({\it cf.} \cite{Romelsberger:2007ec}).
In many $\scN=1$ examples, the theory in the UV is not conformal but flows to a conformal fixed point in
the IR. In these situations, the IR R-symmetry is a mixture of the UV R-symmetry $\tilde r$ and flavor symmetries, and we need to discuss non-trivial anomalous dimensions.
This effect can be incorporated by shifting the flavor chemical
potential $z^F$ by $t^{3(Q-2/3)}=t^{2(3Q/2-1)}$, where $3Q/2$
is the anomalous dimension and the factor $2$
comes from the definition of the index, see \eqref{N1index}.

The total zero-point contribution $\cI^0_{p,m} e^{iB^0_{p,m}(a)}$ from vector and chiral multiplets is
\begin{multline}
\cI^0_{p,m} e^{iB^0_{p,m}(a)} = \exp \left[ \frac{3\beta}{4p} \sum_{\alpha \in G} \big( p \mmod{\alpha(m)} - \mmod{\alpha(m)}^2 \big) \right. \\
\left. - \sum_{\Phi: \,\text{chiral}} \sum_{\rho \in \cR_\Phi} \frac{\beta \big( 3 - 3Q - 2F_\Phi \nu \big) + 2i\rho(a)}{4p} \big( p \mmod{\rho(m)} - \mmod{\rho(m)}^2 \big) \right] \;,
\label{N1I0}
\end{multline}
where the contribution of vanishing weights of the adjoint
representation drops out and $\alpha\in G$ now represents the sum over roots of the gauge
group.

\subsection{Refined Index}

We can construct a refined 4d orbifold index which depends on holonomies for the flavor symmetries, besides the chemical potentials. To construct it, both in $\cN=1$ and $\cN=2$ cases, we first define the flavor chemical potentials in an alternative equivalent way (therefore setting $z=1$ in the previous expressions): we introduce external vector fields for all flavor symmetries. Let $\tilde G = G \times H$ be the extended symmetry group, of which $G$ is gauged and $H$ is external. We do not integrate over the external vector fields in (\ref{N2I}) and (\ref{N1I}), nor introduce a single-letter contribution for them as opposed to (\ref{I2vector}) and (\ref{single letter N=1 vector}). However in the half-hyper (\ref{I_halfhyper}) and the chiral multiplets (\ref{single letter N=1 chiral}) single-letter index, as well in the zero-point energies (\ref{N2I0}) and (\ref{N1I0}), we sum over weights of the representation $\cR$ under the full symmetry group $\tilde G$.

We can introduce holonomies $e^{i a_\alpha}$ of the external vector fields along the temporal direction $S^1_T$: up to conjugation, they are parametrized by parameters $\{a_\alpha\}$ in the maximal torus of $H$. After complexification of the cotangent bundle of the maximal torus, we can identify $e^{ia_\alpha} = z_\alpha$ with the flavor chemical potentials.

For $p>1$ we can also introduce flavor holonomies $e^{2\pi i
m_\alpha/p}$, mutually commuting with the temporal holonomies, along
$S^1_H$ inside $L(p,1)$. The integer parameters $\{m_\alpha\}$, with $0 \leq m_\alpha < p$, provide a refined version of the index:
\be
\cI_p(t,y,v; a_\alpha, m_\alpha) \;.
\ee
Note that the flavor holonomies break the flavor group as $H \to \prod_I H_I$, and enter both in the single-letter indices and in the zero-point energy. On the other hand as we do not integrate over temporal flavor holonomies, we do not sum over flavor holonomies.

The refined index is useful if we want to compute the index of a theory
obtained by gauging together two theories $\cT_1$ and $\cT_2$ along a
common flavor symmetry factor $H'$ (see section \ref{sec.2dTQFT})
\begin{multline}
\cI_{p}(t,y,v; a, m, c, s) = \sum_{r} \cI^{0,\text{ vector }H'}_{p,r} \int \big[ db \big] \, \exp\left[ \sum_n \frac1n \hat \cI^{\text{vector } H'}_{p,r}(t^n,y^n,v^n; e^{inb}) \right] \\
\cI^{\cT_1}_{p}(t,y,v; a,m,b,r) \, \cI^{\cT_2}_{p}(t,y,v; b,r,c,s) \;.
\end{multline}
Here $a,b,c$ are flavor chemical potentials, $m,r,s$ are flavor holonomies, $(b,r)$ refer to the common flavor symmetry $H'$, and the inserted functions are the zero-point energy and single-letter contribution of the gauge fields along $H'$.

In the following sections we discuss the reduction of the 4d orbifold index
to the 3d partition function and the 3d index. Correspondingly, there are
refinements of the 3d index and 3d partition functions. The former is the
generalized superconformal index of \cite{Kapustin:2011jm}.

\section{Relation to the 3d Index on $S^1\times S^2$}
\label{sec.3dindex}

In this section we show explicitly that the $p\to \infty$
limit of the 4d $\scN=1$ index on $S^1\times L(p,1)=S^1 \times S^3/\bZ_p$ gives the 3d
$\scN=2$ index on $S^1 \times S^2$. A parallel analysis shows that in the same
limit the 4d $\scN=2$ index on $S^1 \times L(p,1)$ reduces to the 3d $\scN=4$ index on $S^1 \times S^2$.
This result can be regarded as yet another derivation of the 3d index,
including the non-trivial monopole charges. This approach does not require
more intricate information such as the choice of clever localization
terms and supersymmetry transformation on curved backgrounds.
Notice that the present method can be applied to 3d theories coming from
dimensional reduction from the 4d parents, and cannot be used for
theories with Chern-Simons terms.

Let us take the limit $p\to \infty$. The circle $S^1_H$ shrinks to zero size in this limit and thus the chemical potential
$y$ along the direction goes to $1$.
The expression in the parenthesis appearing in both (\ref{single letter N=1 vector}) and (\ref{single letter N=1 chiral}) gets finite contributions in the limit from either $\mmod{\rho(m)}\sim 0$ or $\mmod{\rho(m)}\sim p$:
\bea
\label{hatIlimit}
\hat\cI^{\scN=1 \textrm{ vector}}_{p,m} &\quad\to\quad  \sum_{\alpha\in G} \Big( - t^{3|\alpha(m)|} + \delta_{\alpha(m),0} \Big) \,  e^{i\alpha(a)} \ , \\
\hat\cI^{\scN=1 \textrm{ chiral}}_{p,m} &\quad\to\quad \sum_{\rho\in \cR} \frac{t^{3Q} z^F  e^{i\rho(a)} -t^{6-3Q} z^{-F}  e^{-i\rho(a)}}{1-t^6}\, t^{3|\rho(m)|} \;.
\eea
Note that the roots with $\alpha(m) = 0$ do not contribute to the vector single-letter index.
 In this limit, the total zero-point contribution $\cI^0_{p,m} e^{iB^0_{p,m}(a)}$ becomes
\beq
\exp\left[ \frac{3\beta}4 \sum_{\alpha\in G} |\alpha(m)| - \sum_{\Phi \text{ chiral}} \sum_{\rho \in \cR_\Phi} \frac{\beta \big( 3 - 3Q - 2 F_\Phi \nu \big) +2i\rho(a) }4 \, |\rho(m)| \right] = t^{3\epsilon_0} z^{q_0} e^{ib_0(a)} \;,
\eeq
with
\bea
\epsilon_0 &= -\frac12 \sum_{\alpha\in G} |\alpha(m)| - \frac12 \sum_{\Phi} \sum_{\rho\in \cR_{\Phi}} \left(Q-1\right) \, |\rho(m)| \ ,\\
q_{0,i} &= -\frac12 \sum_{\Phi} F_i(\Phi) \sum_{\rho\in \cR_{\Phi}} |\rho(m)| \ , \\
b_0(a) &= - \frac12 \sum_\Phi \sum_{\rho \in R_\Phi} \rho(a) \, |\rho(m)| \;.
\eea
In summary, the $p\to \infty$ limit of the 4d index gives rise to
\be
\cI = \sum_{m} t^{3\epsilon_0} z^{q_0} \int [da] \, e^{ib_0(a)} \, \exp \left\{ \sum_{n=1}^\infty \frac1n
\, \hat{\cI}_{p,m}(\cdot^n) \right\} \;,
\ee
where $\hat{\scI}$ is given by the sum of \eqref{hatIlimit}.
This result coincides with the formula for the 3d index given in
\cite{Imamura:2011su},%
\footnote{
With respect to the expression in \cite{Imamura:2011su}, we have $S_\text{CS}^{(0)} = 0$ because our 3d theories arise from the dimensional reduction of 4d theories and
do not have a Chern-Simons term in the Lagrangian.}
\footnote{
\label{flatmeasure}
We could make use of the identity
\be
\prod_{\alpha\in G,\, \alpha(m)=0} 2i \sin \frac{\alpha(a)}2 = \exp\left( \sum_{n=1}^\infty \frac1n \, g(e^{ina}) \right) \qquad\text{with}\qquad g(e^{ia}) = - \sum_{\alpha\in G,\, \alpha(m)=0} e^{i\alpha(a)} \ ,
\ee
to rewrite the Haar measure $[da]$ in terms of the flat measure $\widetilde{[da]}$:
\be
\widetilde{[da]} = \frac1{\prod_I n_I!} \prod_{i=1}^N \frac{da_i}{2\pi} \;,
\ee
reabsorbing the extra factor into the vector multiplet single-letter index. In this case we have
\be
\hat{\cI}_{p,m}^\text{vector, flat} = - \sum_{\alpha \in G} x^{|\alpha(m)|} \, e^{i\alpha(a)} \;,
\ee
which is the expression found in \cite{Imamura:2011su}.
}
provided that we identify $x=t^3$.
Here $x$ appears in the definition of the 3d index
\be
\cI = \Tr \big[ (-1)^\cF x^{(E + j)}  z^{F} \big] \ ,
\label{3dindex}
\ee
and the trace is taken over operators satisfying
\be
\{\cQ,\cQ^\dag\} = E - j - \tilde r =0\ .
\label{3dQQ}
\ee
Comparing \eqref{N1index}, \eqref{N1QQ} and \eqref{3dindex},
\eqref{3dQQ}, we see that $t^{2(E+j_2)} = t^{6j_2 + 3\tilde r}$
should be identified with $x^{E+j} = x^{2j + \tilde r}$.
This explains the parameter identification $x=t^3$.

\section{Relation to the 3d Partition Function on $L(p,1)$}
\label{sec.3dPF}

In this section we consider the 4d $\scN=1, 2$ index on $S_T^1 \times L(p,1)$
and show that in the limit $S_T^1\to 0$ it reproduces the 3d partition
function of the dimensionally reduced 3d $\scN=2, 4$ theory on
$L(p,1)$.%
\footnote{The same problem for $p=1$ was analyzed in
\cite{Dolan:2011rp,Gadde:2011ia,Imamura:2011uw}.
See also \cite{Gang:2009wy,Kallen:2011ny} for the 3d partition function
for a pure gauge theory without matters
on lens spaces and more generally on Seifert manifolds.
Our 3d partition function on a lens space
includes the coupling with matter.}
This is to be expected since
when the circle shrinks the non-trivial modes along the circle become
infinitely massive and decouple from the spectrum,
leaving only the constant modes along $S^1_T$.
Indeed, the Lagrangian of the 4d and
3d theories are the same up to terms irrelevant for the localization
\cite{Imamura:2011uw}, and at the level of the one-loop determinant
the $S^1_T\to 0$ limit is realized as (see \eqref{ZPhi})
\beq
\prod_E \prod_{n=-\infty}^{\infty} \left({\frac{2\pi i
n}{\beta}+E}\right) \to \prod_E E \ ,
\eeq
where we regularized the divergent constant $ \prod_{n\ne 0}\left(\frac{2\pi i
n}{\beta}\right)$.
The right hand side is precisely the one-loop determinant of the 3d
theory.

\subsection{3d Partition Function on $L(p,1)$}

Let us first present the 3d partition function on the lens
space
for general 3d $\scN=2$ theories, in the absence of Chern-Simons terms.
This in itself is a useful result regardless of the reduction from the 4d
index.
The answer can be obtained by generalizing the localization procedure of
 \cite{Kapustin:2009kz,Jafferis:2010un,Hama:2010av}.
The reduction from the 4d index provides another derivation
of this result.

The partition function takes the following form:
\be
Z_{\rm 3d}[L(p,1)] = \sum_m \int  \! \big[ da \big]_{\rm 3d} \,
Z^{\rm vector}_\text{1-loop} [a,m] \,
Z^{\rm chiral}_\text{1-loop} [a,m] \ ,
\label{3dlensePF}
\ee
where $\big[ da \big]_{\rm 3d}$ is a Vandermonde measure of the residual gauge symmetry
\be
\label{3dreducedmeasure}
\big[ da \big]_{\rm 3d}=\frac1{\prod_I N_I!} \prod_{i=1}^N da_i
\prod_{\substack{\alpha \in G \\ \alpha(m) =0}} \alpha(a) \ ,
\ee
and  $Z^{\rm vector}_\text{1-loop}$ and $Z^{\rm chiral}_\text{1-loop}$
are the one-loop determinants of
the gauge and matter sectors. For a vector multiplet:
\be
\label{3dVM1loop}
Z^{\text{vector}}_\text{1-loop} [a,m] = \prod_{\alpha >0} \frac{\sinh \left[\frac{\pi}{p} \big( \alpha(a) + i\alpha (m) \big) \right]
\sinh \left[ \frac{\pi}{p} \big(\alpha (a) - i\alpha (m) \big) \right]}{ \big( \alpha (a) \big)^{2\delta_{\alpha (m), 0}} } \ ,
\ee
whose denominator cancels the Vandermonde measure
\eqref{3dreducedmeasure}.

For a chiral multiplet with an anomalous R-charge $Q$, the one-loop determinant
becomes
\be
\label{3dHM1loop}
Z^{\text{chiral}}_\text{1-loop} [a,m] = \prod_{\rho\in \cR} \prod_{l=0}^\infty \left( \frac{l + 2-Q + i \rho (a)}{l + Q - i \rho (a)}\right)^{N_\rho(l)} \ ,
\ee
where $N_\rho(l)$ is defined to be the number of half-integers
$ m_1 \in \{-\frac{l}{2}, -\frac{l}{2}+1, \dots, \frac{l}{2} - 1, \frac{l}{2}\}$
satisfying
\be
\label{orbifoldmode2}
2 m_1 = \rho (m) \pmod{p} \ .
\ee
The one-loop determinant \eqref{3dHM1loop} becomes trivial for the $\scN=2$ chiral multiplet (which
has $Q=1$) inside
the $\scN=4$ vector multiplet.

\subsection{From the 4d Index to the 3d Partition Function}

Consider the reduction from the 4d $\cN=1$ index to the 3d $\cN=2$ partition function.

Let us start with the chiral multiplet. The orbifold index
given in \eqref{I1halflense} can be written, with the use of \eqref{Fdef} and \eqref{exponentiate},
as
\be
\cI^{\scN=1 \text{ chiral}}_{p,m} =
\prod_{\rho\in \scR} \; \sideset{}{'}\prod_{n_1,n_2 \geq 0 } \;
\frac{ 1-t^{3(n_1+n_2)+6-3Q} y^{n_1-n_2} z^{-F} e^{-i \gamma\rho(a)}}
{1-t^{3(n_1+n_2)+3Q} y^{n_1-n_2} z^{F} e^{i \gamma \rho(a)}} \ ,
\label{prod1}
\ee
where we rescaled $a$ by a factor $\gamma$ for later purpose,
and prime means that the product is over the non-negative integers $n_1, n_2$ satisfying
the orbifold condition $n_1 - n_2 = \rho(m) \pmod{p}$ in \eqref{orbifoldmode}.%
\footnote{This is the infinite product in \eqref{tmpI}, and includes the zero-point contribution.}
The formula above provides a generalization of the elliptic Gamma function.
Let us define an integer $l$ and a half-integer $m_1$ as
\be
\label{lmdef}
l = n_1 + n_2 \;, \qquad m_1 = (n_1-n_2)/2 \ ,
\ee
then the orbifold condition \eqref{orbifoldmode} agrees with that of the 3d partition function \eqref{orbifoldmode2} and we can rewrite
\be
\label{prod2}
\cI^{\scN=1 \text{ chiral}}_{p,m} =
\prod_{\rho\in \scR} \; \sideset{}{'}\prod_{\substack{l \geq 0 \\ m_1 \in \{ - \frac l2, - \frac l2 + 1, \dots , \frac l2 \} }} \;
\frac{ 1-t^{3l+6-3Q} y^{2m_1}  z^{-F} e^{-i \gamma \rho(a)} }
{ 1-t^{3l+3Q} y^{2m_1} z^F e^{i \gamma \rho(a)} } \;.
\ee
Now we set $t=e^{-\gamma/3}$, $y=z=1$ and
take $\gamma \to 0$ limit \cite{Gadde:2011ia}.\footnote{The $\gamma$
here is different from $\beta$ in \eqref{chemical} by a factor
$3/2$, and is the same as the $\beta$ in \cite{Gadde:2011ia}.}
The index then reduces exactly to the 3d partition function of the
chiral multiplet \eqref{3dHM1loop}.

In the same way we can write the 4d index of the vector multiplet \eqref{I1vectorlense} as
\be
\cI^{\scN=1 \text{ vector}}_{p,m} =
\prod_{\alpha\in G}
\left[
\frac{1}{(1-e^{i\gamma\alpha(a)})^{\delta_{\alpha(m),0}}}
\sideset{}{'}\prod_{l,m_1} \,
\frac{ 1-t^{3l} y^{2m_1}  e^{-i \gamma\alpha(a)} }
{ 1-t^{3l+6} y^{2m_1} e^{i \gamma \alpha(a)} } \right] \;.
\ee
As before, setting $y=1$ and taking the $\gamma \to 0$ limit we obtain (up to overall constants independent of the holonomies):
\bea
\label{VMlimit}
& \prod_{\alpha\in G} \left[
\frac{1}{(\alpha(a))^{\delta_{\alpha(m),0}}} \prod_{l=0}^{\infty }\left(  \frac{l + i \alpha
(a)}{l +2+i \alpha (a)}\right)^{N_{\alpha}(l)} \right]
= \prod_{\alpha\in G}
 \left[\frac{1}{(\alpha(a))^{\delta_{\alpha(m),0}}} \prod_{l=0}^\infty \big( l + i \alpha(a) \big)^{N_\alpha(l)-N_\alpha(l-2)} \right] \\
&\quad =
 \prod_{\alpha\in G} \bigg[\frac{1}{ \big( \alpha(a) \big)^{\delta_{\alpha(m),0}}}
\prod_{\substack{ l \ge 0 \\ -l -\alpha(m) \,\in\, p\bZ }} \big( l  + i \alpha(a) \big)
\prod_{\substack{ l\ge 0 \\ l-\alpha(m) \,\in\, p\bZ}} \big( l  + i \alpha(a) \big) \bigg]
\\
&\quad =
\prod_{\alpha>0} \,
\frac{
\sinh\left[ \frac{\pi}{p} \big( \alpha(a)+i\alpha(m) \big) \right]
\sinh\left[ \frac{\pi}{p} \big( \alpha(a)-i\alpha(m) \big) \right]
}
{\big( \alpha (a) \big)^{2 \delta_{\alpha(m), 0}} } \ ,
\eea
where we defined $N_{\alpha}(l)=0$ when $l<0$.
This result coincides with the 3d partition function of the vector multiplet \eqref{3dVM1loop}.
The measure term in the 4d index \eqref{brokenHaarmeasure} becomes that of the
3d partition function \eqref{3dreducedmeasure}
after rescaling $a$ by a factor of $\gamma$, and
this verifies our claim of the relation between the 4d index and the 3d partition function on
the lens spaces.

Finally, let us take a more general limit  $t=e^{-\gamma/3}, y=e^{-\gamma \eta}, z=e^{-\gamma\nu}$ with $\gamma\to 0$ while keeping
$\eta, \nu$ finite. The one-loop determinants
then become:
\bea
\label{3dsquashed1loop}
Z^{\textrm{vector},\eta}_\text{1-loop} [a,m] &=
\prod_{\alpha \in G}  \; \sideset{}{'}\prod_{l,m_1} \,
\frac{ l+2m_1\eta +i \alpha(a) }{ l+2+2m_1 \eta-i \alpha(a) } \\
&=
\prod_{\alpha\in G} \; \sideset{}{'}\prod_{n_1, n_2 \geq 0} \,
\frac{ n_1(1+\eta)+n_2(1-\eta) +i \alpha(a) }{ n_1(1+\eta)+n_2(1-\eta) +2 -i \alpha(a) } \ , \\
Z^{\textrm{chiral}, \eta}_\text{1-loop} [a,m] &=
\prod_{\rho\in \cR} \; \sideset{}{'} \prod_{l,m_1} \,
\frac{l + 2 - Q-\nu F +2m_1 \eta + i \rho (a)}{l+ Q+\nu F +2m_1 \eta - i \rho (a)} \\
&=
\prod_{\rho\in \scR} \; \sideset{}{'}\prod_{n_1, n_2 \geq0 } \,
\frac{ n_1(1+\eta)+n_2(1-\eta) +2-Q-\nu F+i \rho(a) }{ n_1(1+\eta)+n_2(1-\eta) +Q+\nu F -i \rho(a) } \ ,
\eea
where we used \eqref{lmdef}. Notice that these provide generalizations of the hyperbolic hypergeometric Gamma function.
Moreover we see that $\nu\ne 0$ has the effect of changing the anomalous
R-charge $Q$.
The remaining question is the effect of the parameter $\eta$:
when $\nu=0, \eta\ne 0$, the answer coincides with the 3d
partition function on a squashed lens space (generalizing the result of \cite{Hama:2011ea}) with the
squashing parameter $b=\sqrt{\frac{1+\eta}{1-\eta}}$.

\section{Relation to 2d TQFT}\label{sec.2dTQFT}

Let us apply our formalism to the class of 4d $\scN=2$ SCFTs discovered
by Gaiotto \cite{Gaiotto:2009we} (see also \cite{Benini:2009gi, Gaiotto:2009hg, Chacaltana:2010ks}: these are obtained by compactifying the 6d $(2,0)$ $A_{N-1}$ theory on punctured Riemann surfaces $\Sigma$.
In this paper we specialize to the case $N=2$.

To obtain a Lagrangian description, we fix a pants decomposition of the surface.
This is specified by a graph, whose set of internal edges (trivalent
vertices) we denote by $\scG$ ($\scV$). An internal edge $l\in \scG$
corresponds to an $SU(2)$ gauge group, and a trivalent vertex $(l,m,n)\in \scV$ corresponds
to a tri-fundamental hypermultiplet. Since the total gauge group is
$SU(2)^{|\scG|}$, the holonomy is determined by a set of integers
$m^I_l$.

The 4d orbifold index of this theory can be computed to be (recall \eqref{simplevector} and \eqref{simpletri})
\begin{multline}
\scI_{p} = \sum_{m_l^I} \scI^0_{p,m} \int \prod_{l\in \scG} \prod_I
[dU^I_l] \\
\exp \Bigg(  \sum_{n=1}^{\infty} \frac{1}{n} \bigg[\sum_{l\in \scG}f_{p}(\cdot^n)
 \Tr_\text{Adj}(U_l^n)+\sum_{(l,m,n)\in \scV} g_{p}(\cdot^n) \Tr_\text{tri-fund}(U_l^n,
 U_m^n, U_n^n) \bigg] \Bigg) \ ,
\end{multline}
where $f_{p}$ and $g_{p}$ are defined in \eqref{fgdef}.
Let us define
\bea
C_{\alpha_i, \alpha_j, \alpha_k}&\equiv \exp \bigg(
\sum_n \frac{1}{n} g_{p}(\cdot^n) \Tr_\text{tri-fund}(n \alpha_l, n\alpha_m, n
 \alpha_n) \bigg) \ , \\
\eta^{\alpha_k, \alpha_l}&\equiv \exp \bigg( \sum_{n=1}^{\infty} f_{p}(\cdot^n) \Tr_\text{Adj}
 (n\alpha_i) \bigg) \Delta(\alpha)^{-1} \ ,
\eea
where $\Delta(\alpha)$ is the square root of the measure given in \eqref{measureWeyl}.
As in \cite{Gadde:2009kb}, this trivially satisfies the axioms of TQFT, except for
the associativity. We have checked the associativity by series
expansion. The associativity hold not for a fixed holonomy, but after summing over the holonomies.
We conjecture that the associativity holds in general;
this can be regarded as a non-trivial test of the S-duality of 4d $\scN=2$ SCFTs.
It is desirable to give an analytic proof of the associativity.
If this is the case, we have a 2d TQFT whose correlation function
on the Riemann surface coincides with the orbifold index for the 4d
$\scN=2$ theory
characterized by the same Riemann surface.
When the lens space is a three sphere, the 2d theory is proposed to be the $q$-deformed
Yang-Mills theory \cite{Gadde:2011ik}.
To identify the 2d counterpart of the orbifold index, it would be important to understand
its relation to the AGT correspondence \cite{Alday:2009aq}
between 4d $\cN=2$ partition functions on  ALE spaces and 2d Para-Liouville/Toda theories
\cite{Belavin:2011pp,Nishioka:2011jk,Bonelli:2011jx,Belavin:2011tb,Bonelli:2011kv} (see
also \cite{Alday:2009qq, Kimura:2011zf,Matsuura:2005sg}).
In a similar way, one could consider the 6d $\cN=(2,0)$ $A_1$ theory compactified
on a Riemann surface with $\cN=1$ twist, giving rise to 4d $\cN=1$ theories
\cite{Benini:2009mz}.
We expect their superconformal index and orbifold index to describe some 2d
topological theory on the Riemann surface.

\begin{figure}[t]
\begin{center}
\vspace{0.1cm}
\scalebox{0.4}{\input{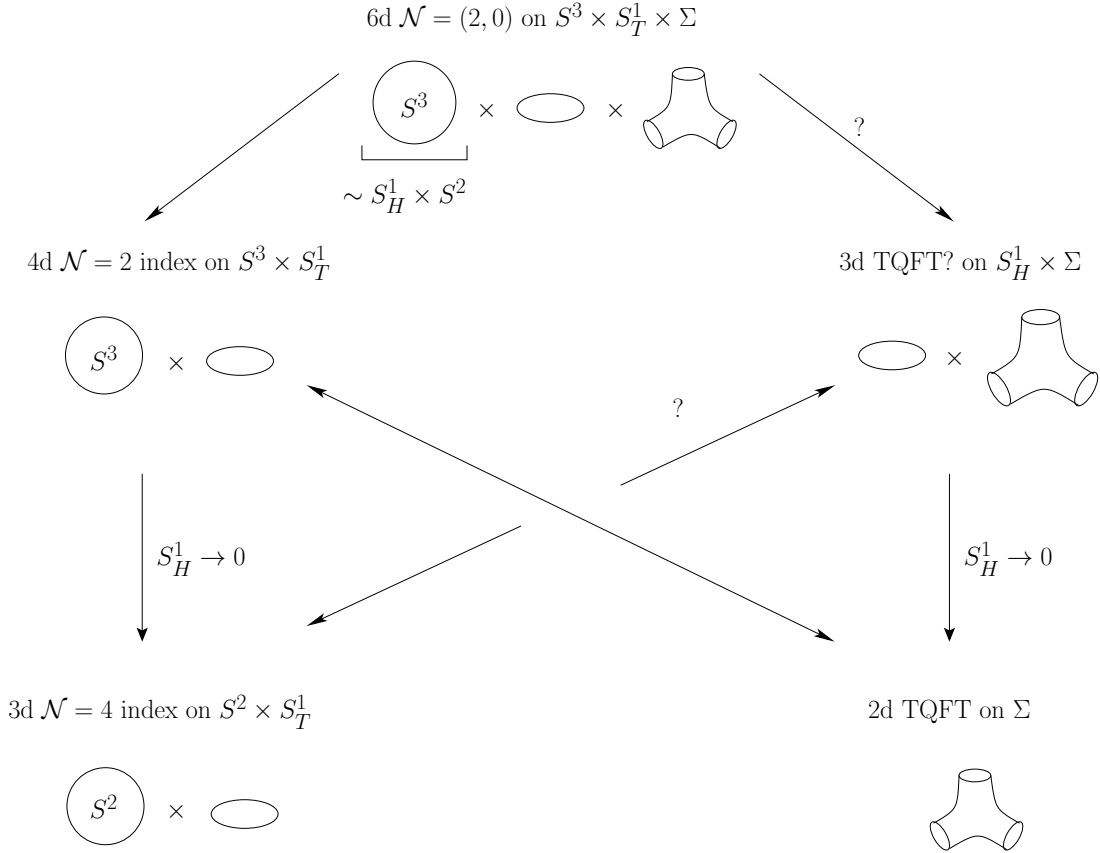}}
\caption{The dimensional reduction from the 4d $\scN=2$ index to the 3d $\scN=4$
index, as discussed in this paper, should correspond to a dimensional
oxidation from the 2d TQFT to a one higher dimensional theory.
}
\label{fig:Relations}
\end{center}
\end{figure}

In the limit $p\to \infty$ the 4d theory reduces to a 3d theory (which by mirror symmetry is dual to a conventional
quiver theory \cite{Benini:2010uu}), the 4d index reduces to the 3d index (section \ref{sec.3dindex})
and it is expected that the
2d TQFT lifts to a 3d TQFT, perhaps along the lines of \cite{ArkaniHamed:2001ca}.
See figure \ref{fig:Relations} for the schematic relations.
Similarly, in the limit $S^1_T\to 0$ ($\gamma\to 0$ in the notation of
section \ref{sec.3dPF}), we expect to recover a Chern-Simons theory with
a non-compact gauge group \cite{Dimofte:2010tz,Terashima:2011qi,Terashima:2011xe,Dimofte:2011jd,Spiridonov:2011hf,Dimofte:2011ju}.
It would be interesting to study these points further.

\section*{Acknowledgments}

We are grateful to Y.\,Imamura, T.\,Kimura, K.\,Ohta and Y.\,Tachikawa
for valuable discussions, and L.F.\,Alday and J.\,Sparks for comments on
an earlier version of this paper.
The work of F.\,B. and T.\,N. was supported in part by the US NSF under Grants No. PHY-0844827
and PHY-0756966.
F.\,B. would like to thank the Weizmann Institute and the Simons Center for Geometry and Physics for hospitality,
and Chiesa di S. Giovanni Battista (Genova) for the beautiful wedding!
M.\,Y. would like to thank PCTS for its support,
and Kavli Institute for Theoretical Physics and Simons Center for Geometry for Physics for hospitality.

\appendix

\section{Derivation of the 4d Index on $S^1\times L(p,1)$}
\label{app: derivation}

In this appendix we derive our orbifold index in the path-integral
formulation ({\it cf.} \cite{Aharony:2003sx}).
Let us deal with the case of the 4d $\scN=2$ SCFT first, and
comment on the $\scN=1$ case later.

The index \eqref{4dindex} is written as
\be
\label{appindex}
\cI = \Tr (-1)^\cF e^{-\beta (E + j_2 + 2\Omega_1 j_1 - \mu(r+R) + \nu F)} \;,
\ee
where we used the notation \eqref{chemical},
and $\cH$ will denote the set of all fields contributing to the index.
The expression \eqref{appindex}
is equivalent to the path-integral over all the fields $\Phi$ in $\cH$
\be
\cI = \int_{\cH} \cD\Phi \, e^{-S[\Phi]} \ ,
\ee
where we impose a periodic boundary condition along $S^1$ for fermions,
and the chemical potentials modify the covariant derivative
with respect
to the Euclidean time, acting on the field $\Phi$ in a representation
$\scR_{\Phi}$, to be
\be
D_0 = \partial_0 - i \rho(a) - j_2 - 2\Omega_1 j_1 + \mu(r+R)-\nu F \ ,
\ee
where $\rho\in \scR_{\Phi}$ stands for the weight of the representation.
Since the index does not depend on gauge couplings, one can perform a
path-integral exactly in the free field limit.
The one-loop contribution from a field $\Phi$ is
\be
Z_{\Phi} = \prod_{\rho \in \cR_\Phi} \Det^{(-1)^{\scF+1}} \left(-D_0^2 +\Delta_{\Phi}\right) \ ,
\ee
where $\Delta_{\Phi}$ is an operator whose eigenvalues are denoted by
$E_{\Phi}^2$.
For example, we have $\Delta_{\Phi}=-\Delta_{S^3}+1$ for a scalar,
where the term $1$ comes from the conformal coupling of the scalar field.

Let us expand the determinant in terms of the eigenvalues of $\partial_0$:
\be
\Det (-D_0^2 + E_{\Phi}^2) = \prod_{n=-\infty}^\infty
\left(\frac{2\pi in}{\beta} + E_{\Phi, +} \right) \left(-\frac{2\pi
in}{\beta} + E_{\Phi, -}\right) \;,
\ee
where
\be
E_{\Phi, \pm} = E_{\Phi} \pm (- i \rho(a) - j_2 - 2\Omega_1 j_1 + \mu(r+R)-\nu F)  \;.
\label{zdef}
\ee
Consequently:
\be
Z_{\Phi}=Z_{\Phi,+} Z_{\Phi,-}\ , \qquad
Z_{\Phi, \pm}=\prod_{E_\Phi} \prod_{n=-\infty}^{\infty}
\left(\frac{2\pi in}{\beta} + E_{\Phi, +} \right) \ .
\label{ZPhi}
\ee
Since $E_{\Phi, +}$ and $E_{\Phi, -}$ share the same energy but opposite charges, we recognize the two terms as the contributions from a particle and its
anti-particle. We therefore have
\be
\cI = \sum_m \int \big[da  \big] \, \prod_{\scH} Z_{\Phi, \pm }{}^{(-1)^{\scF+1}} \ ,
\label{tmpI}
\ee
where the measure $\big[ da \big]$ comes from the gauge fixing
(see \cite{Hikida:2006qb}, section 2.2), and
we summed over the holonomies $m=\{m_i \}$.
The product is over all states contributing to the index.

We can rewrite \eqref{tmpI} with the help of the formula
\be
\prod_{n=-\infty}^\infty (2\pi i n + x) = 2 \sinh \frac{x}{2} = e^{\frac{x}{2}}(1-e^{-x})
= e^{\frac{x}{2}}\exp\left[ -\sum_{m=1}^\infty \frac{1}{m}e^{-mx}\right] \;,
\label{exponentiate}
\ee
which results in (neglecting overall constants)
\be
\cI = \sum_m \int \big[da  \big] \, e^{- \beta\sum_{\cH}
(-1)^\cF\frac{E_{\Phi, \pm}}{2}  } \exp \left[ \sum_{n=1}^\infty \frac{1}{n}
\hat{\scI}(t^n,y^n,v^n,z^n;e^{i n \beta a})\right] \;,
\ee
where we defined the single-letter index $\hat{\scI}$ by
\be
\hat{\scI}(t,y,v,z;e^{i\beta a}) = \sum_{\cH} (-1)^\cF e^{-\beta E_{\Phi,\pm}} \;.
\ee
The exponential factor $ e^{- \beta\sum_{\cH}
(-1)^\cF\frac{E_{\Phi, \pm}}{2}  }$ is the zero-point (Casimir) contribution to the energy and
chemical potentials.
The Casimir part is easily obtained by differentiating the single-letter index with
respect to $\beta$:
\be
\sum_{\cH} (-1)^\cF E =\, -\, \underset{\beta\to 0} {\textrm{Finite}}
\left[\frac{\partial \hat{\scI}}{\partial \beta}\right] \;,
\label{casimircomp}
\ee
where we remove the divergent part in the $\beta\to 0$ limit
\cite{Kim:2009wb,Imamura:2011su}.
We also neglect the holonomy-independent part of the zero-point contribution, since this is merely an overall
shift of the index and does not affect the sum over the holonomies.

\bigskip

The remaining task is to explicitly evaluate the one-loop determinant, or
 equivalently $\hat{\scI}$. This can be carried out by
using the expressions for $\Delta_{\Phi}$ and $E_{\Phi}$. Alternatively,
we can count the operators contributing to the index, see Table \ref{tab: SUSY states}.

\begin{table}[htbp]
$$
\begin{array}{|c||ccccc|c|}
\hline
\text{operators} & E & j_1 & j_2 & R & r & \textrm{contribution to }\hat{\cI} \\
\hline
\hline
\phi & 1 & 0 & 0 & 0 & -1 & t^2v \\
\lambda^1_{\pm} & \frac32 & \pm\frac12 & 0 & \frac12 & - \frac12 & -t^3y,\, -t^3y^{-1} \\
\bar\lambda_{2+} & \frac32 & 0 & \frac12 & \frac12 & \frac12 & -t^4v^{-1} \\
\tilde F_{++} & 2 & 0 & 1 & 0 & 0 & t^6 \\
\partial_{-+} \lambda^1_+ + \partial_{++}\lambda^1_- = 0 & \frac52 & 0 & \frac12 & \frac12 & - \frac12 & t^6 \\
\hline
q & 1 & 0 & 0 & \frac12 & 0 & t^2v^{-1/2} z^F\\
\bar\psi_+ & \frac32 & 0 & \frac12 & 0 & -\frac12 & -t^4v^{1/2} z^{-F}\\
\hline
\partial_{\pm +} & 1 & \pm\frac12 & \frac12 & 0 & 0 & t^3y,\, t^3y^{-1} \\
\hline
\end{array}
$$
\caption{The operators contributing  to the single-letter 4d $\scN=2$
index $\hat{\scI}$. In Euclidean signature, the vector multiplet is given
 by $(\phi, \bar\phi, \lambda^I_\alpha, \bar\lambda_{I\dot\alpha},
 F_{\alpha\beta}, \tilde F_{\dot\alpha\dot\beta})$ while the
 half-hypermultiplet by $(q,\bar q, \psi_\alpha,
 \bar\psi_{\dot\alpha})$.
We also included a
 constraint from the equation of motion for $\lambda^1$.
\label{tab: SUSY states}
}
\end{table}

For a half-hyper multiplet $(q, \psi)$ with a weight $\rho$,
the non-trivial contributions come from
$\partial_{++}^{n_1} \partial_{-+}^{n_2} q$ and
$\partial_{++}^{n_1} \partial_{-+}^{n_2} \bar{\psi}_+$,
where $n_1, n_2$ are non-negative integers.
We therefore have
\beq
\hat{\scI}^{\scN=2 \text{ half-hyper}}=\sum_{\rho\in \scR} \big( t^2 v^{-\frac{1}{2}} z^F e^{i\rho(a)} -
t^4 v^{\frac{1}{2}} z^{-F}  e^{-i\rho(a)} \big) \, \tilde F_{p}(t,y; \rho(m) ) \ ,
\label{Ihalflense2}
\eeq
where we defined
\beq
\label{Fdef}
\tilde F_{p}(t,y; \rho(m) ) = \sideset{}{'}\sum_{n_1,n_2 \geq 0}
\left( t^{3(n_1+n_2)} y^{n_1-n_2} \right) \ ,
\eeq
and the prime in the sum means that we sum over non-negative integers $n_1, n_2$
satisfying the orbifold projection condition \eqref{orbifoldmode}, to be
given shortly.%
\footnote{To compute $\hat{\scI}_\text{orbifold}$ we can first
compute $\hat{\scI}$ for the unorbifolded theory and then impose the
orbifold projection afterwards. This is because the supercharge $\cQ$ used
in the definition of the index commutes with the orbifold action.}
Since the dependence of $\tilde F_p$ on $\rho(m)$ is through (\ref{orbifoldmode}), it clearly only depends on $\rho(m) \pmod{p}$.
Since $\scR$ is a pseudoreal representation, we have $\sum_{\rho\in
\scR}=\sum_{-\rho\in \scR}$ and we can write
\beq
\label{Ihalflense}
\hat{\scI}^{\scN=2\text{ half-hyper}}=\sum_{\rho\in \scR} \big( t^2 v^{-\frac{1}{2}} z^F-
t^4 v^{\frac{1}{2}} z^{-F} \big) \, \tilde F_{p}(t,y; \rho(m)) \, e^{i \rho(a)} \;.
\eeq
The computation for a vector multiplet is similar, except that
the fields $\lambda^1_\pm$ in the Table \ref{tab: SUSY states} require
a special attention since they have non-zero
$2j_1$ charge and come with the constraint of the equation of motion.
The answer is given by%
\footnote{Before taking the orbifold, we have a function
\begin{equation*}
\frac{t^2 v-t^4 v -t^3 (y+y^{-1})+2 t^6}{(1-t^3 y)(1-t^3 y^{-1})}=
\frac{t^2 v-t^4 v+t^6-1}{(1-t^3 y)(1-t^3 y^{-1})}+1 \ .
\end{equation*}
The expression inside the square bracket in \eqref{Ivectorlense}
is the orbifold of this expression.}
\beq
\label{Ivectorlense}
\hat{\scI}^{\scN=2 \text{ vector}} = \sum_{\rho\in \textrm{Adj}}
\Big[ (t^2 v - t^4 v^{-1} +t^6-1) \, \tilde F_{p}(t,y; \rho(m)) + \delta_{\mmod{\rho(m)},0} \Big] e^{i \rho(a)} \;,
\eeq
where we used $\sum_{\rho \,\in\, \textrm{Adj}}=\sum_{-\rho \,\in\, \textrm{Adj}}$ and the function $\mmod{x}$ is defined in (\ref{mmod definition}).

The projection condition is given by
\beq
\label{orbifoldmode}
n_1-n_2 = \rho(m) \pmod{p} \;.
\eeq
To see this, recall that the effect of the holonomy can be locally
removed by a gauge transformation; however, this modifies the global
boundary condition, and we have
a twisted boundary condition.
The integers $n_1$ and $n_2$ are the spins under the phase rotation
of $z_1$ and $z_2$ in \eqref{lenseorbifold}, and \eqref{orbifoldmode}
ensures the single-valuedness of the wavefunction.%
\footnote{The spins $(n_1, n_2)$ are related to the spins $\big( m_1,\frac{l}{2} \big)$ of
$U(1)_1\times U(1)_2\in SU(2)_1\times SU(2)_2$ by a linear
combination, see \eqref{lmdef}.}
Note that the conditions are the same for bosons and
fermions.

We can use \eqref{orbifoldmode} to evaluate the sum in \eqref{Fdef}.
Let us write $n_1 - n_2 = L + kp$, with $L = \mmod{\rho(m)}$ so that $0 \leq L < p$, and $k \in \bZ$.
We divide the sum into $k\ge 0$ and $k<0$, \ie{} (a) $k\in \bZ_{\ge 0}$ and
(b) $\tilde{k}\equiv -k-1 \in \bZ_{\ge 0}$.
Summing over $n_2 , k$ in (a) and $n_1 ,\tilde{k}$ in (b)
we obtain
\be
\tilde F_p(t,y; \rho(m)) = F_p(t,y; \mmod{\rho(m)})
\ee
where $F_p$ is the function given in \eqref{qne0}.

It is straightforward to repeat the analysis for $\scN=1$ theories.
For an $\scN=1$ vector multiplet
\beq
\label{I1vectorlense}
\hat{\scI}^{\scN=1\text{ vector}}=\sum_{\rho\in \textrm{Adj}}
\Big[ (t^6 -1 ) \, \tilde F_{p}(t,y;\rho(m)) + \delta_{\mmod{\rho(m)},0} \Big] e^{i \rho(a)} \ ,
\eeq
and for a chiral multiplet with flavor charge $F$ and anomalous R-charge $Q$,
\beq
\label{I1halflense}
\hat{\scI}^{\scN=1\text{ chiral}}=\sum_{\rho\in \scR} (t^{3Q} z^F e^{i\rho(a)} -
t^{6-3Q}  z^{-F}  e^{-i\rho(a)}) \, \tilde F_{p}(t,y;\rho(m)) \;.
\eeq


\bibliographystyle{utphys}
\bibliography{4d3d_indices}

\providecommand{\href}[2]{#2}\begingroup\raggedright\begin{thebibliography}{10}

\bibitem{Romelsberger:2005eg}
C.~Romelsberger, ``{Counting Chiral Primaries in ${\mathcal{N}}\!=1$, $d=4$
  Superconformal Field Theories},''
  \href{http://dx.doi.org/10.1016/j.nuclphysb.2006.03.037}{{\em Nucl. Phys.}
  {\bfseries B747} (2006) 329--353},
\href{http://arxiv.org/abs/hep-th/0510060}{{\ttfamily arXiv:hep-th/0510060}}.

\bibitem{Kinney:2005ej}
J.~Kinney, J.~M. Maldacena, S.~Minwalla, and S.~Raju, ``{An Index for 4
  Dimensional Super Conformal Theories},''
  \href{http://dx.doi.org/10.1007/s00220-007-0258-7}{{\em Commun. Math. Phys.}
  {\bfseries 275} (2007) 209--254},
\href{http://arxiv.org/abs/hep-th/0510251}{{\ttfamily arXiv:hep-th/0510251}}.

\bibitem{Pestun:2007rz}
V.~Pestun, ``{Localization of Gauge Theory on a Four-Sphere and Supersymmetric
  Wilson Loops},''
\href{http://arxiv.org/abs/0712.2824}{{\ttfamily arXiv:0712.2824 [hep-th]}}.

\bibitem{Kapustin:2009kz}
A.~Kapustin, B.~Willett, and I.~Yaakov, ``{Exact Results for Wilson Loops in
  Superconformal Chern- Simons Theories with Matter},''
  \href{http://dx.doi.org/10.1007/JHEP03(2010)089}{{\em JHEP} {\bfseries 03}
  (2010) 089},
\href{http://arxiv.org/abs/0909.4559}{{\ttfamily arXiv:0909.4559 [hep-th]}}.

\bibitem{Drukker:2010nc}
N.~Drukker, M.~Marino, and P.~Putrov, ``{From weak to strong coupling in ABJM
  theory},'' \href{http://arxiv.org/abs/1007.3837}{{\ttfamily arXiv:1007.3837
  [hep-th]}}.

\bibitem{Herzog:2010hf}
C.~P. Herzog, I.~R. Klebanov, S.~S. Pufu, and T.~Tesileanu, ``{Multi-Matrix
  Models and Tri-Sasaki Einstein Spaces},''
  \href{http://dx.doi.org/10.1103/PhysRevD.83.046001}{{\em Phys.Rev.}
  {\bfseries D83} (2011) 046001},
  \href{http://arxiv.org/abs/1011.5487}{{\ttfamily arXiv:1011.5487 [hep-th]}}.

\bibitem{Jafferis:2010un}
D.~L. Jafferis, ``{The Exact Superconformal R-Symmetry Extremizes Z},''
\href{http://arxiv.org/abs/1012.3210}{{\ttfamily arXiv:1012.3210 [hep-th]}}.

\bibitem{Hama:2010av}
N.~Hama, K.~Hosomichi, and S.~Lee, ``{Notes on SUSY Gauge Theories on
  Three-Sphere},'' \href{http://dx.doi.org/10.1007/JHEP03(2011)127}{{\em JHEP}
  {\bfseries 03} (2011) 127},
\href{http://arxiv.org/abs/1012.3512}{{\ttfamily arXiv:1012.3512 [hep-th]}}.

\bibitem{Bhattacharya:2008zy}
J.~Bhattacharya, S.~Bhattacharyya, S.~Minwalla, and S.~Raju, ``{Indices for
  Superconformal Field Theories in 3,5 and 6 Dimensions},''
  \href{http://dx.doi.org/10.1088/1126-6708/2008/02/064}{{\em JHEP} {\bfseries
  0802} (2008) 064}, \href{http://arxiv.org/abs/0801.1435}{{\ttfamily
  arXiv:0801.1435 [hep-th]}}.

\bibitem{Kim:2009wb}
S.~Kim, ``{The Complete Superconformal Index for ${\mathcal{N}}\!=6$
  Chern-Simons Theory},''
  \href{http://dx.doi.org/10.1016/j.nuclphysb.2009.06.025}{{\em Nucl. Phys.}
  {\bfseries B821} (2009) 241--284},
\href{http://arxiv.org/abs/0903.4172}{{\ttfamily arXiv:0903.4172 [hep-th]}}.

\bibitem{Imamura:2011su}
Y.~Imamura and S.~Yokoyama, ``{Index for Three Dimensional Superconformal Field
  Theories with General R-Charge Assignments},''
  \href{http://dx.doi.org/10.1007/JHEP04(2011)007}{{\em JHEP} {\bfseries 04}
  (2011) 007},
\href{http://arxiv.org/abs/1101.0557}{{\ttfamily arXiv:1101.0557 [hep-th]}}.

\bibitem{Dolan:2011rp}
F.~A.~H. Dolan, V.~P. Spiridonov, and G.~S. Vartanov, ``{From 4D Superconformal
  Indices to 3D Partition Functions},'' {\em Phys. Lett.} {\bfseries B704}
  (2011) 234,
\href{http://arxiv.org/abs/1104.1787}{{\ttfamily arXiv:1104.1787 [hep-th]}}.

\bibitem{Gadde:2011ia}
A.~Gadde and W.~Yan, ``{Reducing the 4D Index to the S$^3$ Partition
  Function},''
\href{http://arxiv.org/abs/1104.2592}{{\ttfamily arXiv:1104.2592 [hep-th]}}.

\bibitem{Imamura:2011uw}
Y.~Imamura, ``{Relation Between the 4D Superconformal Index and the S$^3$
  Partition Function},''
\href{http://arxiv.org/abs/1104.4482}{{\ttfamily arXiv:1104.4482 [hep-th]}}.

\bibitem{Nishioka:2011dq}
T.~Nishioka, Y.~Tachikawa, and M.~Yamazaki, ``{3d Partition Function as Overlap
  of Wavefunctions},'' \href{http://dx.doi.org/10.1007/JHEP08(2011)003}{{\em
  JHEP} {\bfseries 1108} (2011) 003},
  \href{http://arxiv.org/abs/1105.4390}{{\ttfamily arXiv:1105.4390 [hep-th]}}.

\bibitem{Lin:2005nh}
H.~Lin and J.~M. Maldacena, ``{Fivebranes from Gauge Theory},''
  \href{http://dx.doi.org/10.1103/PhysRevD.74.084014}{{\em Phys. Rev.}
  {\bfseries D74} (2006) 084014},
\href{http://arxiv.org/abs/hep-th/0509235}{{\ttfamily arXiv:hep-th/0509235}}.

\bibitem{Hikida:2006qb}
Y.~Hikida, ``{Phase Transitions of Large $N$ Orbifold Gauge Theories},''
  \href{http://dx.doi.org/10.1088/1126-6708/2006/12/042}{{\em JHEP} {\bfseries
  12} (2006) 042},
\href{http://arxiv.org/abs/hep-th/0610119}{{\ttfamily arXiv:hep-th/0610119}}.

\bibitem{Gadde:2009kb}
A.~Gadde, E.~Pomoni, L.~Rastelli, and S.~S. Razamat, ``{S-Duality and 2D
  Topological QFT},'' \href{http://dx.doi.org/10.1007/JHEP03(2010)032}{{\em
  JHEP} {\bfseries 03} (2010) 032},
\href{http://arxiv.org/abs/0910.2225}{{\ttfamily arXiv:0910.2225 [hep-th]}}.

\bibitem{Gadde:2011ik}
A.~Gadde, L.~Rastelli, S.~S. Razamat, and W.~Yan, ``{The 4D Superconformal
  Index from Q-Deformed 2D Yang- Mills},''
\href{http://arxiv.org/abs/1104.3850}{{\ttfamily arXiv:1104.3850 [hep-th]}}.

\bibitem{Shapere:2008zf}
A.~D. Shapere and Y.~Tachikawa, ``{Central charges of N=2 superconformal field
  theories in four dimensions},''
  \href{http://dx.doi.org/10.1088/1126-6708/2008/09/109}{{\em JHEP} {\bfseries
  0809} (2008) 109}, \href{http://arxiv.org/abs/0804.1957}{{\ttfamily
  arXiv:0804.1957 [hep-th]}}.

\bibitem{Romelsberger:2007ec}
C.~Romelsberger, ``{Calculating the Superconformal Index and Seiberg
  Duality},''
\href{http://arxiv.org/abs/0707.3702}{{\ttfamily arXiv:0707.3702 [hep-th]}}.

\bibitem{Kapustin:2011jm}
A.~Kapustin and B.~Willett, ``{Generalized Superconformal Index for Three
  Dimensional Field Theories},''
  \href{http://arxiv.org/abs/1106.2484}{{\ttfamily arXiv:1106.2484 [hep-th]}}.

\bibitem{Gang:2009wy}
D.~Gang, ``{Chern-Simons theory on L(p,q) lens spaces and Localization},''
  \href{http://arxiv.org/abs/0912.4664}{{\ttfamily arXiv:0912.4664 [hep-th]}}.

\bibitem{Kallen:2011ny}
J.~Kallen, ``{Cohomological localization of Chern-Simons theory},''
  \href{http://dx.doi.org/10.1007/JHEP08(2011)008}{{\em JHEP} {\bfseries 1108}
  (2011) 008}, \href{http://arxiv.org/abs/1104.5353}{{\ttfamily arXiv:1104.5353
  [hep-th]}}.

\bibitem{Hama:2011ea}
N.~Hama, K.~Hosomichi, and S.~Lee, ``{SUSY Gauge Theories on Squashed
  Three-Spheres},'' \href{http://dx.doi.org/10.1007/JHEP05(2011)014}{{\em JHEP}
  {\bfseries 1105} (2011) 014},
  \href{http://arxiv.org/abs/1102.4716}{{\ttfamily arXiv:1102.4716 [hep-th]}}.

\bibitem{Gaiotto:2009we}
D.~Gaiotto, ``{${\mathcal{N}}\!=2$ Dualities},''
\href{http://arxiv.org/abs/0904.2715}{{\ttfamily arXiv:0904.2715 [hep-th]}}.

\bibitem{Benini:2009gi}
F.~Benini, S.~Benvenuti, and Y.~Tachikawa, ``{Webs of five-branes and N=2
  superconformal field theories},''
  \href{http://dx.doi.org/10.1088/1126-6708/2009/09/052}{{\em JHEP} {\bfseries
  0909} (2009) 052},
\href{http://arxiv.org/abs/0906.0359}{{\ttfamily arXiv:0906.0359 [hep-th]}}.

\bibitem{Gaiotto:2009hg}
D.~Gaiotto, G.~W. Moore, and A.~Neitzke, ``{Wall-crossing, Hitchin Systems, and
  the WKB Approximation},''
\href{http://arxiv.org/abs/0907.3987}{{\ttfamily arXiv:0907.3987 [hep-th]}}.

\bibitem{Chacaltana:2010ks}
O.~Chacaltana and J.~Distler, ``{Tinkertoys for Gaiotto Duality},''
  \href{http://dx.doi.org/10.1007/JHEP11(2010)099}{{\em JHEP} {\bfseries 1011}
  (2010) 099},
\href{http://arxiv.org/abs/1008.5203}{{\ttfamily arXiv:1008.5203 [hep-th]}}.

\bibitem{Alday:2009aq}
L.~F. Alday, D.~Gaiotto, and Y.~Tachikawa, ``{Liouville Correlation Functions
  from Four-Dimensional Gauge Theories},''
  \href{http://dx.doi.org/10.1007/s11005-010-0369-5}{{\em Lett. Math. Phys.}
  {\bfseries 91} (2010) 167--197},
\href{http://arxiv.org/abs/0906.3219}{{\ttfamily arXiv:0906.3219 [hep-th]}}.

\bibitem{Belavin:2011pp}
V.~Belavin and B.~Feigin, ``{Super Liouville Conformal Blocks from
  ${\mathcal{N}}\!=2$ $SU(2)$ Quiver Gauge Theories},''
\href{http://arxiv.org/abs/1105.5800}{{\ttfamily arXiv:1105.5800 [hep-th]}}.

\bibitem{Nishioka:2011jk}
T.~Nishioka and Y.~Tachikawa, ``{Para-Liouville/Toda Central Charges from
  M5-Branes},''
\href{http://arxiv.org/abs/1106.1172}{{\ttfamily arXiv:1106.1172 [hep-th]}}.

\bibitem{Bonelli:2011jx}
G.~Bonelli, K.~Maruyoshi, and A.~Tanzini, ``{Instantons on ALE Spaces and Super
  Liouville Conformal Field Theories},''
\href{http://arxiv.org/abs/1106.2505}{{\ttfamily arXiv:1106.2505 [hep-th]}}.

\bibitem{Belavin:2011tb}
A.~Belavin, V.~Belavin, and M.~Bershtein, ``{Instantons and 2D Superconformal
  Field Theory},''
\href{http://arxiv.org/abs/1106.4001}{{\ttfamily arXiv:1106.4001 [hep-th]}}.

\bibitem{Bonelli:2011kv}
G.~Bonelli, K.~Maruyoshi, and A.~Tanzini, ``{Gauge Theories on ALE Space and
  Super Liouville Correlation Functions},''
\href{http://arxiv.org/abs/1107.4609}{{\ttfamily arXiv:1107.4609 [hep-th]}}.

\bibitem{Alday:2009qq}
L.~F. Alday, F.~Benini, and Y.~Tachikawa, ``{Liouville/Toda central charges
  from M5-branes},''
  \href{http://dx.doi.org/10.1103/PhysRevLett.105.141601}{{\em Phys.Rev.Lett.}
  {\bfseries 105} (2010) 141601},
\href{http://arxiv.org/abs/0909.4776}{{\ttfamily arXiv:0909.4776 [hep-th]}}.

\bibitem{Kimura:2011zf}
T.~Kimura, ``{Matrix Model from ${\mathcal{N}}\!=2$ Orbifold Partition
  Function},''
\href{http://arxiv.org/abs/1105.6091}{{\ttfamily arXiv:1105.6091 [hep-th]}}.

\bibitem{Matsuura:2005sg}
S.~Matsuura and K.~Ohta, ``{Localization on the D-Brane, Two-Dimensional Gauge
  Theory and Matrix Models},''
  \href{http://dx.doi.org/10.1103/PhysRevD.73.046006}{{\em Phys. Rev.}
  {\bfseries D73} (2006) 046006},
\href{http://arxiv.org/abs/hep-th/0504176}{{\ttfamily arXiv:hep-th/0504176}}.

\bibitem{Benini:2009mz}
F.~Benini, Y.~Tachikawa, and B.~Wecht, ``{Sicilian Gauge Theories and
  ${\mathcal{N}}\!=1$ Dualities},''
  \href{http://dx.doi.org/10.1007/JHEP01(2010)088}{{\em JHEP} {\bfseries 01}
  (2010) 088},
\href{http://arxiv.org/abs/0909.1327}{{\ttfamily arXiv:0909.1327 [hep-th]}}.

\bibitem{Benini:2010uu}
F.~Benini, Y.~Tachikawa, and D.~Xie, ``{Mirrors of 3d Sicilian theories},''
  \href{http://dx.doi.org/10.1007/JHEP09(2010)063}{{\em JHEP} {\bfseries 1009}
  (2010) 063},
\href{http://arxiv.org/abs/1007.0992}{{\ttfamily arXiv:1007.0992 [hep-th]}}.

\bibitem{ArkaniHamed:2001ca}
N.~Arkani-Hamed, A.~G. Cohen, and H.~Georgi, ``{(De)constructing dimensions},''
  \href{http://dx.doi.org/10.1103/PhysRevLett.86.4757}{{\em Phys.Rev.Lett.}
  {\bfseries 86} (2001) 4757--4761},
  \href{http://arxiv.org/abs/hep-th/0104005}{{\ttfamily arXiv:hep-th/0104005
  [hep-th]}}.

\bibitem{Dimofte:2010tz}
T.~Dimofte, S.~Gukov, and L.~Hollands, ``{Vortex Counting and Lagrangian
  3-manifolds},'' \href{http://arxiv.org/abs/1006.0977}{{\ttfamily
  arXiv:1006.0977 [hep-th]}}.

\bibitem{Terashima:2011qi}
Y.~Terashima and M.~Yamazaki, ``{SL(2,R) Chern-Simons, Liouville, and Gauge
  Theory on Duality Walls},'' \href{http://arxiv.org/abs/1103.5748}{{\ttfamily
  arXiv:1103.5748 [hep-th]}}.

\bibitem{Terashima:2011xe}
Y.~Terashima and M.~Yamazaki, ``{Semiclassical Analysis of the 3d/3d
  Relation},'' \href{http://arxiv.org/abs/1106.3066}{{\ttfamily arXiv:1106.3066
  [hep-th]}}.

\bibitem{Dimofte:2011jd}
T.~Dimofte and S.~Gukov, ``{Chern-Simons Theory and S-duality},''
  \href{http://arxiv.org/abs/1106.4550}{{\ttfamily arXiv:1106.4550 [hep-th]}}.

\bibitem{Spiridonov:2011hf}
V.~Spiridonov and G.~Vartanov, ``{Elliptic hypergeometry of supersymmetric
  dualities II. Orthogonal groups, knots, and vortices},''
  \href{http://arxiv.org/abs/1107.5788}{{\ttfamily arXiv:1107.5788 [hep-th]}}.

\bibitem{Dimofte:2011ju}
T.~Dimofte, D.~Gaiotto, and S.~Gukov, ``{Gauge Theories Labelled by
  Three-Manifolds},''
\href{http://arxiv.org/abs/1108.4389}{{\ttfamily arXiv:1108.4389 [hep-th]}}.

\bibitem{Aharony:2003sx}
O.~Aharony, J.~Marsano, S.~Minwalla, K.~Papadodimas, and M.~Van~Raamsdonk,
  ``{The Hagedorn / Deconfinement Phase Transition in Weakly Coupled Large $N$
  Gauge Theories},'' {\em Adv. Theor. Math. Phys.} {\bfseries 8} (2004)
  603--696,
\href{http://arxiv.org/abs/hep-th/0310285}{{\ttfamily arXiv:hep-th/0310285}}.

\end{thebibliography}\endgroup

\end{document}